\documentclass[conference,10pt]{IEEEtran}
\IEEEoverridecommandlockouts
\usepackage{enumitem}
\usepackage{medmath}
\usepackage{array}
\usepackage{booktabs}
\usepackage{tikz}
\usetikzlibrary{
    shapes.geometric,
    arrows.meta,
    positioning,
    calc,
    fit,
    backgrounds
}
\usepackage{url}
\usepackage{titlesec}
\titlespacing*{\section}{0pt}{6pt}{4pt}
\titlespacing*{\subsection}{0pt}{5pt}{3pt}
\titlespacing*{\subsubsection}{0pt}{4pt}{2pt}
\usepackage{enumitem}
\setlist{nosep}

\usetikzlibrary{shapes.geometric, arrows.meta, positioning, fit, backgrounds, shadows}
\usetikzlibrary{positioning,fit,arrows.meta}
\usepackage{tikz}
\usepackage{pgfplots}
\pgfplotsset{compat=newest}
\usepackage{subcaption}
\usepackage{pgfplotstable}
\usepackage{amsmath, amssymb}
\usepackage{xcolor}
\usepackage[font=small,skip=3pt]{caption}

\newtheorem{theorem}{Theorem}[section]

\newtheorem{definition}[theorem]{Definition}

\usepackage{array}
\usepackage{booktabs}
\usepackage{adjustbox}
\usepackage{cite}
\usepackage{amsmath,amssymb,amsfonts}
\usepackage{algorithm}
\usepackage{algorithmic}
\usepackage{graphicx}
\usepackage{textcomp}
\usepackage{xcolor}
\usepackage{tikz}
\usetikzlibrary{positioning} 
\usetikzlibrary{arrows.meta} 
\usetikzlibrary{shapes.geometric} 
\usetikzlibrary{calc} 
\usepackage{titlesec}
\titlespacing*{\section}{1pt}{0.2ex}{0.2ex}
\titlespacing*{\subsection}{1pt}{0.2ex}{0.2ex}
\usepackage{wrapfig}
\usepackage{enumitem}
\setlist[itemize]{leftmargin=1.2em, topsep=0.2ex, itemsep=0.2ex, parsep=0pt}
\setlist[enumerate]{leftmargin=1.2em, topsep=0.2ex, itemsep=0.2ex, parsep=0pt}

\usepackage{setspace}
\usepgfplotslibrary{groupplots}

\def\BibTeX{{\rm B\kern-.05em{\sc i\kern-.025em b}\kern-.08em
    T\kern-.1667em\lower.7ex\hbox{E}\kern-.125emX}}
\begin{document}

\title{X-NegoBox: An Explainable Privacy-Budget Negotiation Framework for Secure Peer-to-Peer Energy Data Exchange}

\author{
Poushali Sengupta, Sabita Maharjan, Frank Eliassen, Yan Zhang*\\
Department of Informatics, University of Oslo, Norway\\ *University of Electronic Science and Technology of China\thanks{\{poushals,sabita,frank\}@ifi.uio.no, yanzhang@ieee.org}
}


\maketitle
\vspace{-1.2em}
\begin{abstract}
The decentralization of modern energy systems is transforming consumers into \textit{prosumers} who continuously exchange data with aggregators, peers, and market operators. While such data is essential for peer-to-peer trading, demand response, and distributed forecasting, it can reveal sensitive household patterns, creating significant privacy risks. Most existing data-sharing mechanisms rely on fixed policies or predetermined differential-privacy budgets, limiting their ability to adapt to variations in reliability, data sensitivity, and intended purpose across requests. As a result, prosumers rarely receive explanations for why a data request is accepted, rejected, or modified, weakening trust and limiting meaningful participation in peer-to-peer data exchange. To address these limitations, we introduce an explainable negotiation framework that enables adaptive privacy budgeting and transparent decision-making. In the proposed framework, each prosumer’s data is managed locally within a \textit{Private DataBox}, where raw data remain confined and are never directly shared. Incoming requests are processed by an \textbf{Autonomous Privacy-Budget Negotiation Protocol (APBNP)}, which dynamically determines an optimal differential-privacy budget by accounting for trust, feature sensitivity, declared purpose, historical sharing behavior, and risk-aware pricing. When a request cannot be accepted as-is, APBNP generates privacy-preserving counter-offers, such as reduced resolution or shorter duration. To ensure transparency, X-NegoBox includes an \textbf{Explainable Agreement Layer (X-Contract)} that produces human- and machine-readable explanations for acceptance, rejection, or modification. After agreement, requester code executes locally inside the DataBox sandbox, and only sanitized outputs are released. All negotiation and optimization steps execute locally within a secure computation environment, while execution authorization is distributed using threshold secret sharing. Experiments on realistic energy-market settings show that X-NegoBox reduces privacy leakage, increases acceptance rates, and provides interpretable justification for decisions.

\end{abstract}

\begin{IEEEkeywords}
Privacy, Differential Privacy, Smart Contracts, Peer-to-Peer Energy Systems, Explainable AI, Data Sharing.
\end{IEEEkeywords}

\section{Introduction}

The reliability and efficiency of decentralized energy systems increasingly
depend on fine-grained data related to electricity consumption, generation,
and flexibility in market operations~\cite{parag2016electricity,zhang2018review}.
While operationally valuable, such data can reveal sensitive household
patterns, including occupancy, appliance usage, and lifestyle
characteristics~\cite{hart1992nonintrusive,molina2010privacy}.\footnote{Examples
include presence inference, appliance-level behavior, and routine detection.}
Existing privacy-preserving approaches, such as static access-control policies,
fixed differential-privacy (DP) budgets~\cite{dwork2006calibrating}, or
centralized clean rooms~\cite{abadi2016deep}, rely on pre-defined configurations
and apply uniform privacy settings across heterogeneous data-access requests,
despite differences in data sensitivity, requester intent, and usage context.
Moreover, prosumers typically receive no clear explanation for why a request
is accepted, rejected, or modified, limiting transparency and trust in
participatory energy systems~\cite{wachter2017counterfactual}.
Advanced metering infrastructure (AMI) and home energy management systems enable
prosumers to participate in peer-to-peer energy exchange and local
balancing~\cite{morstyn2018using}. Such participation requires continuous data
exchange, including load curves and generation
forecasts~\cite{kong2021data}, which can expose sensitive household routines and
socioeconomic attributes~\cite{giani2020smartmeter,beckel2014revealing}.
As data is shared with aggregators, peers, and service providers, concerns
regarding privacy protection and regulatory compliance
intensify~\cite{acquisti2016privacy}. Fine-grained energy data (e.g., smart meter readings) can reveal sensitive household patterns such as occupancy, appliance usage, and behavioral habits, enabling inference attacks and potential misuse (e.g., surveillance or security threats). In peer-to-peer energy trading, repeated data exchange amplifies these risks, motivating adaptive, context-aware privacy mechanisms beyond static policies.

Existing solutions broadly fall into three categories:
\textbf{access control} with static rules~\cite{fernandes2014security},
\textbf{differential privacy} with fixed budgets~\cite{dwork2014algorithmic,ghasemi2021dpenergy},
and \textbf{data clean rooms} that centralize sensitive data~\cite{bonawitz2019towards}.\footnote{These
approaches differ in enforcement and threat models but share static disclosure
assumptions.} However, they fail to accommodate varying privacy preferences across requesters
and purposes~\cite{apthorpe2017spying,gdpr2016}. Supporting peer-to-peer energy
interactions therefore requires an adaptive, bidirectional negotiation mechanism
that dynamically adjusts disclosure parameters (e.g., temporal resolution or
reporting duration), rather than relying on static
policies~\cite{zyskind2015decentralizing,fan2020adaptive}. Without such
mechanisms, prosumers have limited ability to understand or control cumulative
privacy-budget consumption across repeated
requests~\cite{doshi2017towards}. To address these challenges, we introduce \textbf{X-NegoBox}, a distributed,
privacy-preserving framework that enables adaptive control of data disclosure
while keeping all raw data confined within a local \emph{Private DataBox}.
Incoming requests are evaluated by an \emph{Autonomous Privacy-Budget Negotiation
Protocol} (APBNP), which allocates per-request DP budgets based on request
parameters and remaining privacy allowance~\cite{li2016membership}. Requests
operate only on derived data (e.g., aggregates or coarse-grained profiles), while
raw measurements remain local. When a request cannot be satisfied, APBNP
generates privacy-preserving counteroffers by adjusting disclosure parameters.
The contributions of this work are:
\begin{enumerate}[leftmargin=0pt, topsep=0pt]
    \item An \emph{explainable, autonomous} privacy-budget negotiation framework
    for prosumer-to-prosumer energy data exchange.
    \item The \emph{APBNP} protocol for adaptive, per-request DP allocation.
    \item \emph{X-Contract}, an explainability layer providing interpretable
    justification for contract decisions.
    \item A secure local execution sandbox enabling ``run code, not data''
    interactions~\cite{hardy2017private}.
    \item An evaluation on realistic energy-market workloads demonstrating
    improved privacy preservation, trust, and contract acceptance.
\end{enumerate}
The remainder of the paper is organized as follows. Section~II presents
background and motivation. Section~III introduces the system model.
Section~IV details APBNP. Section~V describes X-Contract. Section~VI discusses
sandboxed execution. Section~VII evaluates X-NegoBox, and Section~VIII
concludes the paper.

\section{Stakeholders and Roles}
X-NegoBox operates in a prosumer-controlled environment involving five key
stakeholders. A \textbf{prosumer} is a household or local energy actor that owns
sensitive energy data (e.g., load curves, PV generation, flexibility bids, and
appliance-level signatures) and retains full data ownership. A
\textbf{Private DataBox} is a prosumer-controlled computation enclave deployed at
the household or trusted edge that stores raw data locally, enforces privacy and
contract constraints, and guarantees that raw data never leaves the box. A
\textbf{requester} is an external entity, such as another prosumer, an aggregator,
a forecasting service, or a market operator, seeking access to derived energy data
for a declared purpose (e.g., billing, forecasting, or peer-to-peer trading). A
\textbf{contract agent} is a lightweight orchestration component within the
DataBox that validates requests, coordinates the Autonomous Privacy-Budget
Negotiation Protocol (APBNP), generates counter-offers when required, and invokes
the explainable agreement layer (X-Contract). Finally, an
\textbf{execution sandbox} provides an isolated runtime for executing
requester-supplied code under strict contract and privacy constraints, injecting
differential-privacy noise and releasing only sanitized outputs.
\begin{figure}[htbp]
\centering
\small
\begin{adjustbox}{width=\linewidth}
\begin{tikzpicture}[
    node distance=1 cm and 2cm,
    every node/.style={font=\sffamily\large, align=center},
    base/.style={draw, rounded corners=8pt, minimum width=5cm, minimum height=4.5cm, line width=1.1pt},
    external/.style={base, fill=blue!10, draw=blue!60!black},
    internal/.style={base, fill=gray!5, draw=gray!70},
    logic/.style={base, fill=orange!10, draw=orange!80!black},
    explain/.style={base, fill=green!10, draw=green!70!black},
    sandbox/.style={base, fill=red!10, draw=red!80, dashed, line width=1.5pt},
    data/.style={circle, draw=gray!70, fill=gray!20, minimum size= 1.4cm, line width=1.2pt},
    arrow/.style={->, >=stealth, thick, draw=gray!80},
    dasharrow/.style={->, >=stealth, dashed, thick, draw=gray!60, shorten >=3pt, shorten <=3pt},
    dataarrow/.style={->, >=stealth, very thick, draw=red!60, line width=1.4pt},
    rejectarrow/.style={->, >=stealth, thick, draw=red!70}
]

\node[external] (requester) {
    \textbf{Requester}\\
    \Large (Aggregator or Prosumer)
};

\draw[fill=gray!6, draw=gray!50, rounded corners=15pt, line width=1.8pt] 
    (4, -20) rectangle (17.5, 2.5);
\node[anchor=north east, font=\sffamily\bfseries\LARGE\color{gray!80}] at (13.3, 2.2) {PRIVATE DATABOX};

\node[internal, right=5cm of requester] (contract) {
    \textbf{\LARGE 1. Contract Agent}\\\\
    \LARGE  Validates \& Parses Proposal\\
    \textit{\LARGE Checks syntax, intent,}\\
    \textit{\LARGE  and resource needs}
};

\node[logic, below=1.2cm of contract] (apbnp) {
    \textbf{\LARGE 2. APBNP}\\\\
    \textit{\LARGE Negotiates $\epsilon$, $\delta$ }
};

\node[explain, below=1.5cm of apbnp] (xcontract) {
    \textbf{\LARGE 3. X-Contract}\\\\
    \LARGE Explainable Decision Layer
};
\node[sandbox, below=1.2cm of xcontract] (sandbox) {
    \textbf{\LARGE 4. Execution Sandbox}\\ \\
    \LARGE Isolated code execution\\
    \LARGE + Differential Privacy noise injection\\
    \textit{\LARGE Runs requester-supplied code safely}
};

\node[external, right=4cm of xcontract] (output) {
    \textbf{\LARGE Sanitized Output}\\
    \LARGE Differentially Private Results\\
    \LARGE ($\epsilon$, $\delta$)-DP guaranteed
};

\node[data, left=4.7cm of sandbox] (rawdata) {
    \textbf{RAW}\\
    \textbf{DATA}
};

\draw[arrow] (requester) -- node[above, font=\LARGE] {1. Contract Proposal\\(Code + Query\\ + Proposed $\epsilon$)} (contract);

\draw[arrow] (contract) -- node[left, font=\LARGE, pos=0.6] {2. Validated Proposal} (apbnp);

\draw[arrow] (apbnp) -- node[right, font=\LARGE, pos=0.4] {3. Allocated Privacy Budget\\Accepted / Rejected} (xcontract);

\draw[arrow] (xcontract) -- node[left, font=\LARGE, pos=0.6] {4. Approved Code + Budget} (sandbox);

\draw[arrow] (sandbox.east) -- ++(3.2,0) node[midway, below, font=\LARGE] {5. Noisy/Aggregated \\ Results} |- (output.south);

\draw[very thick, draw=red!70] (sandbox.east) -- ++(5.5,0) node[pos=0.6, above right, font=\LARGE\bfseries, text=red!80!black] {NO RAW DATA EVER LEAVES};

\node[below=0.5cm of apbnp.south, font=\LARGE, xshift=-3.8cm, text=red!70] (reject) {Rejection + Counter-proposal};
\draw[rejectarrow] (apbnp.south) -- ++(0,-1) -| ([xshift=-0.5cm]requester.south) 
    node[pos=0.25, below, font=\LARGE] {Reject or Renegotiate};

\draw[dasharrow] (xcontract.west) -- ++(-3,0) 
    node[left, font=\LARGE, align=right] {Household Feedback\\(Accept / Modify / Reject)};

\draw[dasharrow] (xcontract.east) -- node[below, font=\LARGE] {Natural Language \\Explanation \& \\ Reasoning Trace} (output.west);

\draw[dataarrow] (rawdata) -- node[below, sloped, font=\LARGE\bfseries, text=red!70] {Private Raw Data\\(Accessible ONLY \\inside sandbox)} (sandbox.west);

\end{tikzpicture}
\end{adjustbox}
\caption{X-NegoBox architecture enforcing a secure \emph{``run code, not data''} model, where contracts are negotiated, explained, and executed locally with differential privacy, and only sanitized outputs are released.}
\label{fig:architecture-detailed}
\end{figure}
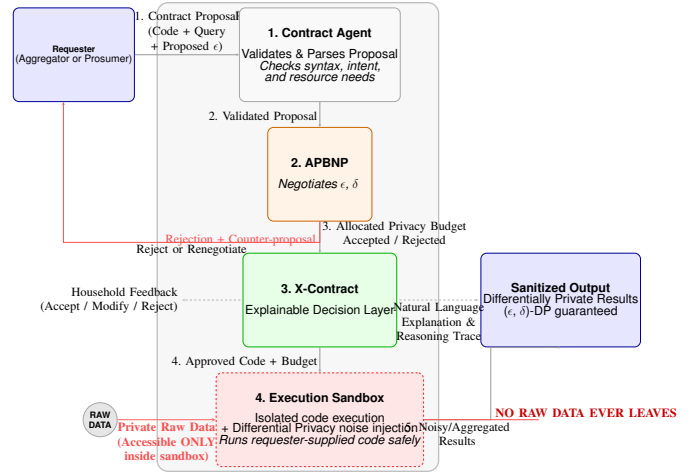

Figure~\ref{fig:architecture-detailed} provides a high-level view of how
data-access requests are processed within X-NegoBox, from the initial contract
proposal to privacy-budget allocation and decision explanation, without exposing
raw household data. The workflow proceeds as follows:
\begin{enumerate}[leftmargin=*]
    \item \textbf{Contract Proposal:}
    A requester submits a contract specifying data fields, temporal granularity,
    duration, intended purpose, maximum acceptable noise, and expected utility.\footnote{These
    parameters define both the disclosure scope and the acceptable privacy--utility
    trade-off from the requester’s perspective.}

    \item \textbf{APBNP Decision Process:}
    Upon receiving a data-access request, the DataBox evaluates request-specific
    parameters such as the requested data granularity, temporal resolution,
    reporting duration, and the remaining privacy allowance. Based on this
    evaluation, the Autonomous Privacy-Budget Negotiation Protocol (APBNP)
    determines an appropriate differential-privacy budget $\varepsilon^\star$
    and selects one of three outcomes: approval, modification, or denial. APBNP proceeds in six phases: (i) authenticate and validate the encrypted request, (ii) compute sensitivity, trust, and purpose scores, (iii) optimize the privacy
budget $\varepsilon^\star$, (iv) check feasibility constraints, (v) generate a
counter-offer or rejection if needed, and (vi) authorize release via threshold
key reconstruction followed by differentially private execution. Approval
    occurs if the request fits within the remaining privacy budget; modification
    produces a counter-offer by adjusting disclosure parameters (e.g., reduced
    temporal resolution or duration); denial is issued when no feasible
    configuration satisfies privacy constraints..

    \item \textbf{X-Contract Explanation:}
    The system produces a transparent explanation detailing the decision
    rationale, privacy--utility trade-offs, and actionable guidance for revising
    the request. An \emph{Explainable Agreement Layer} (X-Contract) generates
    human- and machine-readable justifications for each decision~\cite{doshi2017towards}.
    After agreement, the requester code executes locally in a secure sandbox,
    ensuring that only DP-sanitized outputs are released~\cite{costan2016intel}.

    \item \textbf{Local Execution:}
    After agreement, requester code is validated and executed in the DataBox
    sandbox. Only contract-compliant, DP-sanitized outputs are returned.
\end{enumerate}

\textbf{Privacy Objectives and Operational Safeguards: }
The X-NegoBox architecture is designed with privacy as a primary operational objective, supported by safeguards that regulate how data-access requests are evaluated, adapted, and executed within the DataBox. Rather than exposing raw data, the system limits disclosure to privacy-preserving outputs and enforces constraints throughout the request lifecycle. Data-access requests may involve repeated interactions over time, which can lead to unintended information leakage if not properly controlled. For example, attempts to infer sensitive household attributes, such as occupancy or appliance usage, from released outputs are mitigated through adaptive privacy-budget allocation and controlled output granularity. To prevent reconstruction of exact values across successive queries, the system monitors cumulative privacy loss and dynamically adjusts or restricts further disclosure when necessary. 
\begin{wraptable}{r}{0.58\linewidth}
\centering
\caption{Adaptive X-Contract parameterization across energy data requests.}
\label{tab:xcontract-adaptation}
\setlength{\tabcolsep}{3pt}
\renewcommand{\arraystretch}{0.9}
\begin{adjustbox}{width=\linewidth}
\begin{tabular}{lcccc}
\toprule
\textbf{Param.} &
\textbf{Base} &
\textbf{Settlement} &
\textbf{Forecast} &
\textbf{DR} \\
\midrule
Data type
& HH load
& Agg. load
& Smoothed agg.
& Load + flex. \\

Resolution
& 15\,min
& 1\,h
& 30\,min
& 5\,min \\

Window
& 30\,d (hist.)
& 60\,d (hist.)
& 14\,d roll.
& Event \\

DP $\varepsilon$
& 4.0
& 2.0
& 3.0
& 5.5 \\

Output
& Aggregates
& Billing
& Stats
& Events \\

Mode
& Periodic
& One-shot
& Periodic
& On-demand \\

Filters
& --
& --
& --
& Peaks \\
\bottomrule
\end{tabular}
\end{adjustbox}
\end{wraptable} All data processing is performed within a secure local sandbox that isolates execution from external access. The sandbox validates submitted code, restricts system operations, and prevents unauthorized data extraction, thereby ensuring that privacy guarantees are consistently enforced. The X-NegoBox architecture focuses on five key goals: 
\begin{itemize}[leftmargin=0pt, topsep=0pt]
    \item \textbf{Preserving privacy} by ensuring data sharing meets differential-privacy standards.
    \item \textbf{Automating privacy-budget negotiations} for ease of use.
    \item \textbf{Providing clear explanations} for contract decisions to build trust.
    \item \textbf{Maintaining logs for accountability} and dispute resolution.
    \item \textbf{Ensuring that requesters follow contractual guidelines} for data outputs.
\end{itemize}
Together, these elements, including APBNP, X-Contract, and the local execution sandbox, facilitate a \textit{"run code, not data"} model, enabling secure and transparent data exchanges among prosumers while laying the groundwork for negotiation logic and explainability.
\section{Autonomous Privacy-Budget Negotiation Protocol (APBNP)}
APBNP determines whether a contract is approved, rejected, or modified by
computing an optimal privacy budget $\varepsilon^\star$ that balances privacy
risk, expected utility, trust, sensitivity, purpose, and historical budget
consumption. This section formalizes the notation, optimization criteria,
scoring components, and negotiation algorithm. Let requester $i$ and data owner
$j$ be prosumers. A contract specifies a feature set $x$, time window $w$,
temporal resolution $r$, and purpose $p$. The sensitivity of $x$ is denoted by
$S_x$, trust by $T_{ij}$, and purpose compatibility by $P(p)$. Let $H_j$ denote
prosumer $j$’s cumulative privacy-budget consumption. Functions $U(\varepsilon)$
and $R(\varepsilon)$ represent the expected utility and privacy risk under
budget $\varepsilon$, respectively, while $C(\varepsilon)$ captures the
associated economic cost. The resulting optimal budget is denoted
$\varepsilon^\star$.
To regulate the privacy budget offered to a requester, X--NegoBox first quantifies 
the inherent privacy risk associated with the features included in the request.  
We formalize this using a \emph{sensitivity score}, which aggregates feature-level 
risk weights into a single scalar.

\begin{definition}[\textbf{Feature Sensitivity Weight}]
Let $f$ denote a feature category included in a data-access request $x$.  
Each feature is assigned a sensitivity coefficient $\alpha_f \in [0,1]$ that 
reflects its inherent privacy risk. Larger values indicate a higher likelihood 
of revealing personal, behavioural, or appliance-level patterns.
\end{definition}
Typical values used throughout this work are: $\alpha_{\text{loc}} = 1.0,
\alpha_{\text{app}} = 0.7,
\alpha_{\text{load}} = 0.4, 
\alpha_{\text{agg}} = 0.2.$
\begin{definition}[\textbf{Sensitivity Score}]
For a request $x$ containing a set of features $\mathcal{F}(x)$, the sensitivity 
score is defined as
\begin{equation}
\medmath{
S_x \;=\; \sum_{f \in \mathcal{F}(x)} \alpha_f.
}
\label{eq:sensitivity-score}
\end{equation}
\end{definition}

A higher $S_x$ indicates that the request targets more privacy-critical aspects of 
the household profile.Consequently, a higher sensitivity score reduces the set of privacy budgets $\varepsilon$ that can be safely allocated to a request. Rather than directly rejecting such requests, the negotiation engine adapts the contract parameters to remain within acceptable privacy limits. Concretely, this adaptation may involve lowering the temporal or spatial granularity of the data, increasing the amount of noise injected into the released outputs, or restructuring the request into an alternative form that exposes less sensitive information. Sensitivity scoring therefore serves as an initial control mechanism that converts high-level semantic descriptions of requested data (e.g., appliance-level usage or location-related features) into explicit quantitative constraints that guide privacy-budget allocation and contract reformulation.


Beyond the intrinsic sensitivity of the requested features, the willingness of a 
data owner to share information also depends on the historical relationship with 
the requester. We model this with a contract-level \emph{trust score}, which aggregates 
behavioural evidence accumulated across previous interactions.

\begin{definition}[\textbf{Trust Score}]
Let $i$ denote a requester and $j$ the data owner.  
The trust score $T_{ij}$ quantifies the historical reliability of requester $i$ 
with respect to owner $j$ and is computed as,
\begin{equation}
\medmath{
T_{ij}
    = \beta_1 N_{ij}^{\text{succ}}
    + \beta_2 Q_{ij}^{\text{qual}}
    + \beta_3 A_{ij}^{\text{match}},
}
\label{eq:trust-score}
\end{equation}
where each component captures a distinct behavioural signal.
\end{definition}

The terms in Eq.~\eqref{eq:trust-score} are defined as follows:
$N_{ij}^{\text{succ}}$: number of successfully completed contracts between requester $i$
and owner $j$, indicating behavioral consistency.
$Q_{ij}^{\text{qual}}$: a metric evaluating the compliance, correctness, and timeliness
of outputs from requester $i$.
$A_{ij}^{\text{match}}$ measures how closely actual data use aligns with the
declared purpose, penalizing misuse. Higher trust increases the data owner’s
confidence and allows a wider feasible $\varepsilon$-range, enabling more
informative counter-offers when prior cooperation is strong. Conversely, low
trust reduces the maximum permissible $\varepsilon$ and may trigger fallback
policies such as reporting limitations or additional verification. The
sensitivity score $S_x$ and trust score $T_{ij}$ jointly govern negotiation by
capturing complementary aspects of a request: sensitivity characterizes
\textit{what} information is requested, while trust captures \textit{who} is
requesting it. Sensitivity reflects intrinsic data risk (e.g.,
granularity or temporal scope), whereas trust reflects requester behavior and
historical compliance. In peer-to-peer energy trading, households are generally willing to share
fine-grained load data with trusted neighbors with whom they have previously
traded, while restricting disclosure to aggregated data for new or infrequent
partners. This reflects common practice in community energy markets,
where repeated successful interactions increase willingness to share operational
details. Such history may permit a higher privacy budget $\varepsilon$, but this
remains bounded by data sensitivity and the remaining cumulative budget; even
with strong compliance records, highly sensitive requests remain tightly
constrained to prevent long-term privacy loss. While related to decentralized reputation systems, the proposed mechanism
operates differently.\footnote{Traditional reputation systems typically make
binary allow/deny decisions.} Instead, it employs a continuous reliability score
in $[0,1]$, derived from historical adherence, output quality, and successful
completions with greater weight on recent behavior. Higher scores allow more
informative disclosure, whereas lower scores enforce stricter aggregation and
noise, supporting cooperation while preserving long-term privacy guarantees.

The declared purpose of a request plays a central role in determining how strictly 
privacy should be enforced. Certain purposes (e.g., billing) are legally mandated 
and operationally essential, whereas others (e.g., behavioural profiling) carry 
substantial privacy risks. To formalize this effect, we define a \emph{purpose 
compatibility score} as follows.

\begin{definition}[\textbf{Purpose Compatibility}]
Let $p$ denote the declared purpose of a data request.  
The compatibility function, $\medmath{P(p) \in [0,1]}$ quantifies the legitimacy and societal acceptability of the purpose. 
Higher $P(p)$ values correspond to purposes that align with regulatory norms and 
grid-operational requirements.
\end{definition}
\par 
In practice, the value of $P(p)$ is determined using a predefined compatibility table 
derived from regulatory guidelines, standard energy-market operations, and commonly 
accepted industry practices. Purposes that are explicitly required for settlement, 
billing, or grid operation are assigned higher scores, while purposes that are optional, 
exploratory, or weakly regulated receive lower values. This mapping is fixed, auditable, 
and independent of the requester.
Representative examples used throughout this work include 
$P(\text{billing}) = 1.0$, 
$P(\text{forecasting}) = 0.8$, 
$P(\text{demand response}) = 0.7$, 
$P(\text{peer trading}) = 0.6$, and 
$P(\text{profiling}) = 0.1$. 
Higher compatibility values correspond to purposes that justify lower noise
injection or broader access, while low-scoring purposes trigger stronger noise
or counter-offers with reduced granularity. Thus, $P(p)$ acts as a normative
filter in the negotiation process, complementing sensitivity $S_x$ and
interaction reliability $T_{ij}$. In this work, $P(p)$ is static and assigned via
a predefined lookup table based on the operational necessity of the declared
purpose. Core market functions (e.g., billing or settlement) receive the highest
scores; widely used but non-mandatory activities (e.g., forecasting or grid
monitoring) receive slightly lower scores; and optional or exploratory purposes
(e.g., behavioural profiling) receive low scores due to limited necessity and
higher privacy impact. Each purpose is mapped to a fixed value in $[0,1]$ prior to
deployment and applied uniformly across requests, ensuring transparency,
auditability, and resistance to strategic purpose rephrasing. Representative
categories and scores are shown in Table~\ref{tab:purpose-compatibility}.



\subsection{Privacy-Budget Optimization and Decision Logic}
\label{subsec:apbnp-optimization}
The Adaptive Privacy-Budget Negotiation Protocol (APBNP) seeks to find the optimal privacy parameter, denoted as $\varepsilon^\star$, by balancing four key factors in responsible data sharing: the usefulness of the data, the privacy risk based on feature sensitivity and remaining budget, the trust between the requester and data owner, and the legitimacy of the request's purpose. The protocol assesses each potential $\varepsilon$ value by evaluating a combined score that rewards utility, trust, and legitimate purpose, while penalizing privacy risks and associated long-term costs. The goal is to maximize this score to achieve the best compromise between data utility and privacy protection. The optimal value is the one that maximizes $\varepsilon$ is, 
\begin{equation}
\resizebox{\columnwidth}{!}{$
\varepsilon^\star =
\arg\max_{\varepsilon \in [0, \varepsilon_{\max}]}
\Big(
\lambda_1 U(\varepsilon)
- \lambda_2 R(\varepsilon \mid S_x, H_j)
+ \lambda_3 T_{ij}
+ \lambda_4 P(p)
- \lambda_5 C(\varepsilon)
\Big)
$}
\label{eq:opt-epsilon}
\end{equation}
APBNP selects the privacy budget $\varepsilon$ that maximizes data utility while
limiting privacy risk, accounting for declared purpose, trust, and cost, with
non-negative coefficients $\lambda_k$ weighting their relative influence.\footnote{The
weights $\lambda_k$ regulate the trade-off between utility and privacy
preservation.} Solving Eq.~\eqref{eq:opt-epsilon} yields one of three outcomes:
(i) \emph{approval}, when $\varepsilon^\star$ satisfies all feasibility
constraints; (ii) \emph{rejection}, when no feasible $\varepsilon$ meets minimum
privacy guarantees; or (iii) a \emph{counter-offer}, in which adjusted parameters
$(w,r,x)$ produce a policy-compliant $\varepsilon^\star$. In all cases, APBNP
verifies that $\varepsilon^\star$ does not exceed the data owner’s remaining
cumulative privacy budget, thereby preventing excessive information leakage.
\begin{definition}[\textbf{Feasibility Constraints}]
A privacy budget $\varepsilon^\star$ is feasible if it satisfies:
\begin{equation}
\medmath{\varepsilon^\star \ge \varepsilon_{\min}(S_x, p),}\
\medmath{\varepsilon^\star \le H_j^{\mathrm{remaining}},}
\label{eq:decision-constraints}
\end{equation}
\end{definition}
where $\varepsilon_{\min}$ is the minimum noise level required to protect the
requested feature set given its sensitivity $S_x$ and declared purpose $p$.
If both inequalities hold, the contract is approved and execution proceeds to
secure key reconstruction and differentially private release.  
Otherwise, APBNP attempts to generate a counter-offer by adjusting the
reporting window, temporal resolution, or requested feature granularity.  
If no such modification yields a feasible $\varepsilon$, the system issues a
formal rejection to maintain strict privacy guarantees.
\subsection{Secure-Computation \& TSS Integration}
\label{subsec:secure-env}
The X--NegoBox negotiation engine operates entirely inside a dedicated
\emph{secure computation environment} (SCE), ensuring confidentiality,
integrity, and correctness of all negotiation artefacts, including contract
metadata, feature-sensitivity scores, trust evidence, purpose compatibility
evaluations, and intermediate computations used to determine the optimal
privacy budget $\varepsilon^\star$. The SCE follows a privacy-by-design
architecture combining authenticated encryption, isolated execution, and a
threshold secret sharing (TSS) mechanism for key management. All contract
requests are authenticated using the requester’s public key, and metadata such as load for (e.g., purpose and features) is protected via authenticated encryption
$\text{Enc}_{K^{\text{pub}}}(p,x,r)$, preventing inspection or tampering by
intermediate nodes.Only the Secure Contract Executor inside the SCE holds the private keys, and all
negotiation and decision logic executes locally within the prosumer-controlled
\begin{wraptable}{l}{0.58\linewidth}
\centering
\caption{Purpose-compatibility scores $P(p)$ (higher = more operationally essential).}
\label{tab:purpose-compatibility}
\setlength{\tabcolsep}{3pt}
\renewcommand{\arraystretch}{0.9}
\begin{adjustbox}{width=\linewidth}
\begin{tabular}{p{3.2cm} p{5cm} c}
\toprule
\textbf{Purpose} & \textbf{Rationale} & $\boldsymbol{P(p)}$ \\
\midrule
Billing / Settlement
& Billing, market settlement, regulatory compliance
& 1.0 \\
Load Forecasting
& Short- and medium-term demand prediction
& 0.8 \\
Grid Monitoring
& Congestion, voltage, and anomaly detection
& 0.75 \\
P2P Energy Trading
& Optional, user-driven local energy exchange
& 0.6 \\
Behavioural Profiling
& Habit inference; not operationally required
& 0.1 \\
\bottomrule
\end{tabular}
\end{adjustbox}
\end{wraptable}
DataBox. This includes APBNP optimization, feasibility checks, purpose
compatibility assessment, and explanation generation, all performed using local
metadata and predefined policy tables, without any centralized coordinator,
shared global state, or external computation. As a result, neither raw household
data nor intermediate negotiation artefacts leave the local execution boundary. TSS operates at a separate authorization layer and is
invoked only after contract approval to authorize the release of differentially
private outputs. The signing key is split across independent entities (e.g.,
trusted enclaves or organizationally distinct edge services), such that in a
$(k,n)$ configuration (e.g., $(3,5)$), at least $k$ nodes must cooperatively
reconstruct the key. This separation keeps sensitive decision-making local to the
data owner while distributing execution authorization to prevent unilateral or
premature disclosure, allowing X--NegoBox to combine local optimization and
explainability with distributed authorization.

\begin{figure}[htbp]
\centering
\begin{adjustbox}{width=\linewidth}
\begin{tikzpicture}[
    >=Stealth, thick,
    local/.style={
        fill=green!12,
        draw=green!60!black,
        rounded corners,
        minimum width=10cm,
        minimum height=6cm
    },
    tss/.style={
        fill=orange!18,
        draw=orange!70!black,
        dashed,
        rounded corners,
        minimum width=2.4cm,
        minimum height=1.2cm,
        align=center
    },
    process/.style={
        rectangle,
        draw,
        rounded corners=3pt,
        minimum height=2.6em,
        minimum width=5.2em,
        align=center,
        fill=gray!8
    }
]

\node[local] (sce) {};
\node[font=\bfseries, yshift=-5mm] at (sce.north)
{Secure Computation Environment (prosumer-controlled)};

\node[process] (input)  at (-2.6, 1.9) {Input \& Auth};
\node[process] (policy) at (-0.6, 0.9) {Policy \& Trust};
\node[process] (optim)  at (1.4, -0.1) {$\varepsilon^\star$};
\node[process, fill=red!12] (dp) at (-0.6, -1.9) {DP Release};

\node[tss] (tss1) at (4.6, 1.1) {TSS Node 1\\Key share $s_1$};
\node[tss] (tss2) at (4.6, 0.0) {TSS Node 2\\Key share $s_2$};
\node[tss] (tss3) at (4.6,-1.1) {TSS Node 3\\Key share $s_3$};

\node[font=\small, above=0.1cm of tss1]
{$\geq k$ shares required};

\draw[->] (input) -- (policy);
\draw[->] (policy) -- (optim);
\draw[->] (optim) -- (dp);

\draw[->, densely dashed, red!70!black]
(tss1.west) -- ++(-0.8,0) |- (dp.east);
\draw[->, densely dashed, red!70!black]
(tss2.west) -- ++(-0.8,0) |- (dp.east);
\draw[->, densely dashed, red!70!black]
(tss3.west) -- ++(-0.8,0) |- (dp.east);

\node[left=2cm of input] (req) {Requester};
\draw[->] (req.east) -- node[above, font=\scriptsize]
{encrypted request} (input.west);

\node[right=3.9cm of dp] (out) {Consumer};
\draw[->] (dp.east) -- node[above, font=\scriptsize]
{$\widetilde{D}_j$} (out.west);

\end{tikzpicture}
\end{adjustbox}
\caption{Adaptive Privacy-Budget Negotiation Protocol.
Negotiation runs locally in the prosumer-controlled Secure Computation
Environment; Threshold Secret Sharing (TSS) authorizes a one-time
differentially private release, preventing unilateral disclosure.}
\label{fig:apbnp-tss-visual}
\end{figure}
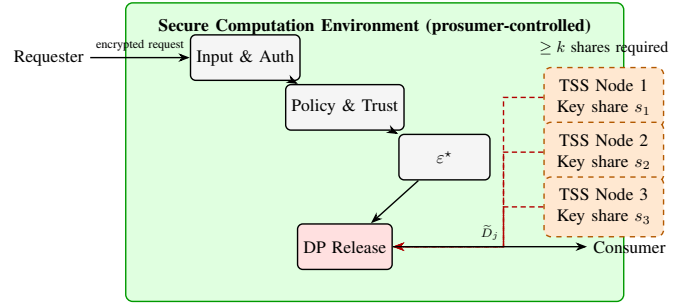

\begin{definition}[\textbf{Threshold Secret Sharing}]
A $(k,n)$ TSS scheme splits a cryptographic secret $s$ into $n$ shares $s \longrightarrow (s_1, s_2, \dots, s_n),$
such that any subset of at least $k$ shares can reconstruct $s$, while any subset of size $<k$ yields zero information about $s$.
\end{definition}
Typical deployments use $(k,n)=(3,5)$ or $(4,7)$, balancing resilience to node
failure with compromise resistance. In Threshold Secret Sharing (TSS), a signing
key is split into $n$ shares and distributed across independent nodes such that
any subset of at least $k$ nodes can reconstruct the key, while fewer than $k$
reveal no information. This prevents any single entity from authorizing data
release without majority agreement. TSS is activated once a contract is ready for execution. After a safe privacy
budget $\varepsilon^\star$ is determined, a reconstruction request is issued, and
the key is recovered only if sufficient trusted nodes cooperate:
$\medmath{\{s_i\}_{i \in \mathcal{K}} \Longrightarrow \text{Reconstruct}(s),
\|\mathcal{K}| \ge k.}
$ The reconstructed secret $s$ is used exactly once to authorize the
privacy-preserving transformation
$\medmath{D_j \xrightarrow{\text{DP with } \varepsilon^\star} \widetilde{D}_j,}
$after which the key is immediately erased from memory. Even an adversary with temporary control over one or two nodes cannot force data
release or weaken DP noise parameters. Authorization is possible only if
(i) the negotiation protocol determines a safe $\varepsilon^\star$, and
(ii) the threshold of trusted nodes jointly reconstructs the key.
We employ established TSS constructions (e.g., Shamir’s polynomial-based scheme
or modern lattice-based variants) rather than introducing a new protocol.
Designing a next-generation, negotiation-aware TSS that embeds policy constraints
directly into the reconstruction process is left for future work. Figure~\ref{fig:apbnp-tss-visual} illustrates the resulting authorization workflow
in X-NegoBox, showing how negotiation and explainability remain local within a
Trusted Execution Environment, while execution authorization is distributed
across independent threshold nodes.


The computational overhead of APBNP is modest and well suited for deployment at
the edge, such as within a household DataBox or trusted gateway device. The
privacy-budget selection problem is formulated as a one-dimensional optimization
over the interval $\varepsilon \in [0, \varepsilon_{\max}]$, evaluated in practice
over a finite set of $M$ candidate values, resulting in $O(M)$ time
complexity.\footnote{In practice, $M$ is small and fixed by design (e.g., tens of
candidate values), yielding negligible latency.}
The remaining components of the negotiation are lightweight. Sensitivity scoring
and application-level compatibility evaluation operate only on request metadata








 \begin{wrapfigure}{r}{0.73\linewidth}
\begin{minipage}{\linewidth}
\begin{algorithm}[H]
\scriptsize
\caption{X-Contract Explanation}
\label{alg:xcontract}
\begin{algorithmic}[1]
\REQUIRE Decision $d$, scores $(S_x, T_{ij}, P(p), H_j^{\text{rem}})$, optimized $\varepsilon^\star$
\IF{$d = \text{approve}$}
  \RETURN ApprovalExplanation($S_x, T_{ij}, P(p), \varepsilon^\star, U_{\text{PU}}$)
\ELSIF{$d = \text{reject}$}
  \RETURN RejectionExplanation($S_x, H_j^{\text{rem}}$)
\ELSIF{$d = \text{counter}$}
  \RETURN CounterOfferExplanation($w, r, \varepsilon^\star$)
\ENDIF
\end{algorithmic}
\end{algorithm}
\end{minipage}
\end{wrapfigure}and scale linearly with the number of requested features, which is typically
small.\footnote{These operations are independent of dataset size or historical
record length.}
Feasibility checks involve simple comparisons against precomputed thresholds and
the remaining privacy budget, and therefore execute in constant time. The threshold secret sharing (TSS) mechanism is invoked only after a contract has
been approved and does not participate in the optimization process itself.
Key reconstruction requires cooperation among $k$ out of $n$ key-holding
entities, incurring $O(k)$ communication and computation cost, where $k \ll n$ in typical deployments.\footnote{Reconstruction occurs at most once per approved
contract, so its overhead is amortized and does not affect the negotiation
phase.} Overall, the latency of APBNP is dominated by local metadata processing and simple
arithmetic operations. As a result, negotiation time remains predictable,
bounded, and independent of the volume of raw data stored in the DataBox, making
the framework practical for real-time and resource-constrained edge environments.

\section{X-Contract: Explainable Agreement Layer}

While APBNP determines the optimal privacy budget $\varepsilon^\star$ and feasibility of a contract, prosumers require understandable justifications to make informed decisions about data sharing. To address this, X-NegoBox incorporates \textbf{X-Contract}, an explainable agreement layer that generates transparent, human-readable and machine-interpretable explanations for every negotiation outcome. X-Contract increases prosumer trust, improves accountability, and supports regulatory compliance in energy data ecosystems. X-Contract is built on three principles: \textbf{Transparency}: Prosumers must understand why a contract is approved, rejected, or counter-offered. \textbf{Actionability}: Explanations should provide concrete steps (e.g., reduce time window) that improve the likelihood of approval. \textbf{Traceability}: Every explanation is logged, ensuring accountability and enabling auditing of privacy decisions. For each negotiation outcome produced by APBNP, X-Contract provides one of three explanation types.
\begin{itemize}[leftmargin=0 pt, topsep=0pt, itemsep=0pt]
    \item \textbf{Approval Explanation}: Clarifies why the request satisfies privacy constraints.
    \item \textbf{Rejection Explanation}: Identifies which constraints were violated and why no feasible $\varepsilon^\star$ exists.
    \item \textbf{Counter-Offer Explanation}: Suggests modifications to contract parameters to achieve a valid privacy budget.
\end{itemize}
These explanations rely on the underlying factors computed by APBNP: sensitivity, trust, purpose, and historical budget. X-Contract summarizes the tradeoff between privacy and utility through the privacy--utility score:
\begin{equation}
   \medmath{ U_{\text{PU}} = \lambda_1 \big(1 - R(\varepsilon^\star)\big)
              + \lambda_2 U(\varepsilon^\star)
              + \lambda_3 T_{ij}
              - \lambda_4 C(\varepsilon^\star).}
\end{equation}
A higher value indicates that the contract provides strong utility with acceptable risk.  
The score is included in the explanation to inform both parties. X-Contract relies on a rule engine mapping APBNP outputs to explanations. Let $d \in \{\text{approve}, \text{reject}, \text{counter}\}$ be the decision.  
The explanation logic is:
\begin{equation}
    \medmath{\text{Explain}(d) = 
\begin{cases}
\text{Exp}_{\text{app}}(T_{ij}, S_x, P(p), \varepsilon^\star) & d = \text{approve}, \\
\text{Exp}_{\text{rej}}(S_x, H_j^{\text{rem}}) & d = \text{reject}, \\
\text{Exp}_{\text{ctr}}(w, r, \varepsilon^\star) & d = \text{counter}. \\
\end{cases}}
\end{equation}
X-Contract explains why a data-sharing request is approved, rejected, or
modified by translating the outcome of APBNP into a human-readable rationale.
For approved requests, explanations highlight the key factors enabling safe
acceptance, such as acceptable sensitivity, sufficient remaining privacy
budget, and appropriate declared use. For rejected requests, the explanation
explicitly identifies the violated constraint (e.g., excessive sensitivity or
insufficient remaining budget). When a counteroffer is issued, X-Contract
clarifies which parameters were adjusted, such as reduced temporal resolution
or increased noise, and why the modified request becomes acceptable. Each
explanation is generated from template-based functions referencing computed
scores, thresholds, the privacy–utility score $U_u$, and actionable guidance
for revising future requests. We illustrate this with a peer-to-peer energy trading scenario in which
Prosumer~A requests 5-minute flexibility data from Prosumer~B for one day.
With sufficient remaining budget ($H_j=5.0$), the request is approved as
low-sensitivity and allocated $\varepsilon^\star=2.25$. With lower but feasible
budget ($H_j=1.5$), the request is approved with $\varepsilon^\star=1.50$ and an
explicit warning that most remaining privacy capacity is consumed. When the
budget is insufficient ($H_j=0.3$), the request is rejected to prevent privacy
risk, and the requester is guided to revise the request by reducing resolution,
duration, or feature sensitivity. These cases show that X-Contract delivers
consistent, auditable explanations aligned with the underlying negotiation
logic. X-Contract provides transparent and interpretable decision reasoning that is
largely absent from existing privacy-preserving systems. By exposing actionable
feedback and scoring factors, it enables prosumers to understand negotiation
outcomes. Algorithm~\ref{alg:xcontract} summarizes the explanation-generation
process for approval, rejection, and counter-offer decisions. Robustness is
evaluated by perturbing interaction-reliability and sensitivity scores by
$\pm5\%$. Across $2000$ replayed requests, explanation outcomes remain unchanged
in $97.25\%$ of cases, with deviations occurring only at feasibility threshold
crossings. This indicates that explanations are locally stable and aligned with
the underlying negotiation logic rather than numerical noise.
\section{Local Execution Sandbox}

After a contract is approved or adjusted via APBNP and X-Contract, the requester
submits computation code for execution within the prosumer’s Private DataBox.
Since raw data never leaves prosumer control, the execution sandbox enforces
contract compliance, differential-privacy guarantees, and safe execution of
untrusted code. Prior to execution, the DataBox validates that the code accesses
only the approved feature set $x$, time window $w$, and granularity $r$; uses
only whitelisted operations (e.g., arithmetic, aggregation, inference); passes
lightweight static analysis for prohibited file or network access, recursion, or
exfiltration logic; and executes under a bounded runtime to prevent
denial-of-service attacks. Validation failures are explained via X-Contract. Validated code executes in an isolated sandbox (e.g., containerized VM,
lightweight enclave, or WebAssembly runtime) with no access to system files,
external processes, or network interfaces. Memory safety is enforced by bounding
heap and stack usage, and execution tracing is used to detect suspicious
behavior. After computing the output $y$, the DataBox applies differential
privacy noise based on the negotiated budget $\varepsilon^\star$, producing
$\tilde{y}=y+\text{Noise}(\varepsilon^\star,S_x)$, where noise magnitude
increases with sensitivity $S_x$ and decreases with higher trust or purpose
compatibility. Supported mechanisms include Laplace noise for numerical
aggregates, Gaussian noise for model outputs, and randomized rounding for
categorical values; privacy loss and noise application are logged by
X-Contract. Finally, the DataBox enforces contract compliance by restricting output
dimensionality, capping resolution, and rejecting outputs that expose raw values
or exceed the DP budget. Violations halt execution and are logged. Through
validation, isolation, and DP enforcement, the sandbox mitigates repeated-query,
model-inversion, side-channel, and code-injection attacks, operationalizing the
\emph{run code, not data} paradigm while ensuring privacy, integrity, and
explainability throughout the data-access lifecycle.

\section{Experimental Evaluation}
\label{sec:experiments}
In this section we evaluate the \textbf{X-NegoBox} framework and the \textbf{APBNP}
negotiation engine. All experiments are implemented in Python and executed on a
local workstation. Performance is assessed along three axes: (i) a
\emph{48-scenario cross-dataset benchmark} using real load time series,
proxy-real traces derived from real data, and synthetic loads; (ii) a
\emph{full-scale synthetic smart-grid simulation} with 100 heterogeneous
prosumers; and (iii) a \emph{privacy-budget sweep} varying the initial privacy
budget. These experiments evaluate privacy preservation, negotiation stability,
acceptance behavior, and sensitivity to budget availability. All experiments
reuse identical request streams and real datasets to ensure controlled and
comparable evaluation, without relying on synthetic fallback data.

\textbf{Datasets:} For the analysis, we utilize two dataset families: real/proxy-real datasets like the \textbf{UCI Household Dataset} and national-level load data from \textbf{DE, FR, IT (Energy-Charts API)}, along with synthetic city-level datasets generated for \textbf{Oslo, Berlin, Rome, and Paris}. We also create a \textbf{synthetic prosumer ecosystem} with 100 prosumers and 60 days of hourly data. This ecosystem features varied load profiles influenced by factors such as temperature-dependent heating and appliance usage, enabling the exploration of appliance-level sensitivities, trust distribution, and privacy-budget dynamics across multiple interactions. We simulate a trust-inflation adversary that behaves cooperatively to
increase its trust score over time.
Despite trust saturating at its maximum value, cumulative privacy
leakage remains bounded by the remaining budget.
In the experiment, the adversary obtains $49$ accepted contracts, after
which the remaining budget drops below $0.05$.
This confirms that trust alone cannot override cumulative budget and
sensitivity constraints in \textsc{APBNP}.

\textbf{Assumptions and Execution Environment:}
Our evaluation assumes: (i) a secure local execution environment with encrypted storage and authenticated communication; (ii) trusted DataBox modules maintaining $H_j^{\text{max}}$, history, and local DP mechanisms; (iii) confidentiality of contract metadata, privacy budgets, and TSS keys; and (iv) local DP noise injection before any data leaves the DataBox. X-NegoBox targets resource-constrained edge environments (e.g., smart meters), where APBNP achieves linear-time complexity with millisecond latency, and communication is minimized via “run code, not data.” The system assumes a semi-trusted decentralized setting with SCE enforcement, TSS authorization ($<k$ of $n$ compromised), and trust-aware but constrained negotiation. While realistic, deployment may require lightweight cryptography, batching, and hardware enclaves, and assumptions should be revisited under adversarial or large-scale conditions. Under this setting, APBNP enables adaptive negotiation of $\varepsilon^\star$ based on trust, sensitivity, purpose, and cost while preserving privacy constraints.

\textbf{Negotiation Model and Privacy Mechanisms:}
Each prosumer $j$ maintains an individual privacy budget $H_j$, representing the
remaining allowable privacy loss.
In the synthetic ecosystem, budgets are initialized as
$H_j^{(0)} = 8.0,\ \forall j \in \{1,\dots,100\}$, and are decremented as contracts
are accepted. Given a request defined by a purpose $p$ (e.g., forecasting, billing), a feature
set $F$ (e.g., hourly load, PV generation, EV events), a requester trust score
$T \in [0,1]$, and a feature sensitivity score $S \ge 0$, the APBNP engine searches
for a feasible privacy parameter $\varepsilon^\star$ by maximizing the following
scoring function:
\begin{equation}
\label{eq:score_function}
\text{Score}\medmath{(\varepsilon) =
    2\sqrt{\varepsilon}
    - 1.8 \cdot S \varepsilon^{1.7}
    + 1.0T + 0.8P
    - 0.15 \varepsilon.}
\end{equation}
Here, $\medmath{S=\sum_{f\in F}\text{sensitivity}(f)}$ and $\medmath{P=\text{purpose-score}(p)}$.
The term $2\sqrt{\varepsilon}$ models utility gains from reduced noise, while
$-1.8\,S\,\varepsilon^{1.7}$ penalizes privacy risk for sensitive features.
A regularizer $-0.15\,\varepsilon$ discourages excessive budgets, and $T$ and $P$
encode trust and purpose awareness. The optimizer searches for
$\varepsilon^\star$ over $0<\varepsilon\le\varepsilon_{\max}$, subject to
$\varepsilon^\star \le H_j$, ensuring that allocated privacy loss does not exceed
the remaining budget. If no feasible $\varepsilon^\star$ exists under the
initial sensitivity $S$, X-NegoBox applies a three-stage fallback.
\begin{wraptable}{r}{0.48\linewidth}
\centering
\caption{Mean acceptance rates across datasets and negotiation patterns.}
\label{tab:acceptance-comparison}
\setlength{\tabcolsep}{4pt}      
\renewcommand{\arraystretch}{0.9} 
\tiny
\begin{tabular}{lccc}
\toprule
\textbf{Dataset} & \textbf{Calibrated} & \textbf{Fixed} & \textbf{Natural} \\
\midrule
Berlin & 0.599 & 0.597 & 0.593 \\
DE     & 0.604 & 0.609 & 0.593 \\
FR     & 0.608 & 0.595 & 0.589 \\
IT     & 0.595 & 0.603 & 0.598 \\
Oslo   & 0.613 & 0.599 & 0.585 \\
Paris  & 0.588 & 0.609 & 0.605 \\
Rome   & 0.594 & 0.603 & 0.604 \\
UCI    & 0.592 & 0.590 & 0.599 \\
\midrule
\textbf{Mean} & \textbf{0.599} & \textbf{0.601} & \textbf{0.596} \\
\bottomrule
\end{tabular}
\end{wraptable}
First, it issues a counter-offer by reducing sensitivity to $0.25S$ (via
downsampling, aggregation, or feature removal) and re-optimizes.
If unsuccessful, it may allow a minimal $\varepsilon$ only for highly trusted
requesters in low-risk scenarios.
If both strategies fail, or if the safety condition $H_j < 4S$ holds, the request
is rejected to prioritize privacy protection.
Upon approval, the allocated $\varepsilon^\star$ is deducted from $H_j$,
enforcing cumulative privacy-risk accounting for future negotiations.

\textbf{Baseline Comparison:} We compare \textsc{APBNP} against a fixed-$\epsilon$
differential privacy baseline using identical request streams on the same real
datasets. Evaluation focuses on acceptance rate, cumulative privacy leakage, and
budget exhaustion time. Across all datasets, \textsc{APBNP} achieves a mean
acceptance rate of $0.9998$ while keeping cumulative privacy leakage within the
initial budget ($H^{(0)}=8.0$) and avoiding premature exhaustion. In contrast,
the fixed-$\epsilon$ baseline collapses: for $\epsilon_{\text{fix}}=0.10$,
acceptance drops to $0.04$, and the budget is exhausted after only $80$ requests.
These results show that static privacy policies cannot sustain continuous
operation under realistic workloads, whereas \textsc{APBNP} supports long-term
data sharing through adaptive budget shaping rather than rejection.
\subsection{Experimental Design}
\label{subsec:experimental_design}
We measure the runtime overhead of \textsc{APBNP} negotiation and
\textsc{X-Contract} explanation generation on a commodity CPU.
Privacy-budget optimization takes $0.32$\,ms per request, while explanation
generation adds $0.006$\,ms. Sandboxed execution incurs a $121\%$ relative
overhead, but absolute execution time remains below $0.02$\,ms, indicating that
\textsc{X-NegoBox} is suitable for real-time edge deployment. Robustness to
parameter mis-specification is assessed by perturbing scoring weights, trust
distributions, and feature-sensitivity coefficients by $\pm10\%$. We conduct
three complementary experiments:
\begin{itemize}[leftmargin=0pt]
    \item \textbf{Experiment 1: Cross-Dataset Benchmark (48 scenarios).}
    We simulate 2000 requester--owner interactions across eight datasets.
    Each interaction specifies a request purpose, feature sensitivities, and a
    dataset-specific trust score. Three negotiation patterns
    (fixed-acceptance, natural, dataset-calibrated) are evaluated under two risk
    modes (mild, moderate), yielding 48 scenarios to assess acceptance stability
    and policy robustness.
    \item \textbf{Experiment 2: Full Synthetic Smart-Grid Simulation.}
    Using the corrected APBNP engine, we model a synthetic ecosystem of 100
    prosumers and simulate 2500 interactions. Requests follow realistic purpose
    and sensitivity profiles (utilities, aggregators, third-party services). We
    analyze large-scale privacy-budget evolution, rejection versus fallback
    behavior, and feasibility of the $\varepsilon^\star$ scoring function under
    heterogeneous trust and sensitivity conditions.
    \item \textbf{Experiment 3: Privacy-Budget Sweep.}
    For each initial privacy budget $\varepsilon_0 \in \{1,\ldots,10\}$, we
    simulate 700 interactions using fixed trust, purpose, and sensitivity
    distributions. Acceptance, rejection, and counter-offer rates are measured to
    characterize the transition from scarce-privacy (tight budgets) to
    surplus-privacy (high feasibility) regimes.
\end{itemize}
Across all variants, acceptance rates vary within a narrow band of width
$0.001$, indicating stable negotiation behavior and low sensitivity to moderate
hyperparameter variation.
\subsection{Results and Discussion}
\label{subsec:results}
\textbf{Cross-Dataset Acceptance Behavior:} Across all 48 cross-dataset experiments, we observe that the acceptance rate lies in a narrow, highly stable band: $\medmath{0.57 \le \text{Acceptance} \le 0.62.}$ Table~\ref{tab:acceptance-comparison}  visualizes this as a heatmap over datasets (rows) and negotiation patterns/risk modes (columns).  It also highlights structural differences 
between household and national load series, demonstrating the heterogeneity under which X-NegoBox is evaluated. The global mean and standard deviation across all scenarios are:$ \bar{A} = 0.603, \sigma = 0.013.$ Table~\ref{tab:acceptance-comparison} reports the mean acceptance rates aggregated per dataset. Two observations stand out:

\begin{enumerate}[leftmargin=0pt]
    \item \textbf{Lack of dataset-induced collapse or inflation.} Despite differences in scale (household vs national), climate, and synthetic generation, all acceptance rates concentrate around $\approx 0.60$. This indicates that the APBNP scoring function and constraints are robust to distributional changes in load time series.
    \item \textbf{Balanced negotiation behavior.} Acceptance around $0.60$ reflects a regime where requests are neither trivially accepted nor systematically rejected. X-NegoBox thus avoids ``all-or-nothing'' behavior, which is a known issue in simpler differential-privacy contracts.
\end{enumerate}

In practical terms, this means that operators can deploy X-NegoBox across different regions and datasets without having to retune the entire negotiation logic for each scenario.
\begin{figure}[htbp]
\centering
\begin{tikzpicture}
\begin{axis}[
    width=0.85\linewidth,
    height=3.2cm,
    xmin=1, xmax=10,
    ymin=0, ymax=1.05,
    xtick={1,2,3,4,5,6,7,8,9,10},
    ytick={0,0.2,0.4,0.6,0.8,1.0},
    xlabel={Initial Privacy Budget $\epsilon_0$},
    ylabel={Decision Rate},
    legend style={at={(0.5,-0.25)}, anchor=north, legend columns=3},
    grid=both,
    grid style={dashed,gray!20},
]

\addplot[
    thick,
    blue,
    mark=*,
] coordinates {
    (1,0.55) (2,0.74) (3,0.85) (4,0.97) (5,0.99)
    (6,0.99) (7,0.99) (8,1.00) (9,1.00) (10,1.00)
};

\addplot[
    thick,
    red,
    mark=square*,
] coordinates {
    (1,0.40) (2,0.24) (3,0.14) (4,0.03) (5,0.01)
    (6,0.01) (7,0.01) (8,0.00) (9,0.00) (10,0.00)
};

\addplot[
    thick,
    orange,
    mark=triangle*,
] coordinates {
    (1,0.05) (2,0.02) (3,0.01) (4,0.00) (5,0.00)
    (6,0.00) (7,0.00) (8,0.00) (9,0.00) (10,0.00)
};

\legend{Acceptance, Rejection, Counter-offer}

\end{axis}
\end{tikzpicture}

\caption{APBNP decision rates as a function of the initial privacy budget $\epsilon_0$.  
Acceptance rises monotonically, rejection decreases, and counter-offers vanish beyond $\epsilon_0 \ge 4$.}
\label{fig:sweep_plot}
\end{figure}
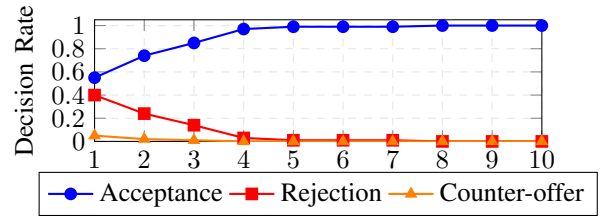

\noindent\textbf{Full Synthetic Prosumers (100-Household Simulation:}) In the 100-prosumer synthetic experiment involving 2500 interactions, the system achieved an acceptance rate of 87.49\%, with rejection at 12.51\% and zero counter-offers. The mean remaining privacy budget per prosumer was 0.765, indicating that the negotiation engine effectively produces feasible $\varepsilon^\star$ 
\begin{wrapfigure}{r}{0.75\linewidth}
\centering
\begin{tikzpicture}
\begin{axis}[
    width=0.9\linewidth,
    height=3.7cm,
    title={Acceptance Rate Across 48 Experiments.},
    xlabel={Experiment Index},
    ylabel={Acceptance Rate},
    xmin=-1, xmax=48,
    ymin=0.57, ymax=0.705,
    xtick={0,5,10,15,20,25,30,35,40,45},
    ytick={0.58,0.60,0.62,0.64,0.66,0.68,0.70},
    grid=major,
    grid style={gray!30,dashed},
    legend style={
        at={(0.98,0.55)},
        anchor=south east,
        font=\small,
        draw=none
    },
    tick label style={font=\small},
    label style={font=\small},
    title style={font=\small},
]

\addplot[
    color=green!60!black,
    mark=*,
    mark size=2.5pt,
    thick
] coordinates {
(0,0.577) (1,0.602) (2,0.583) (3,0.616) (4,0.598) (5,0.586)
(6,0.614) (7,0.603) (8,0.598) (9,0.589) (10,0.620) (11,0.588)
(12,0.606) (13,0.583) (14,0.605) (15,0.573) (16,0.616) (17,0.601)
(18,0.607) (19,0.599) (20,0.594) (21,0.603) (22,0.606) (23,0.585)
(24,0.602) (25,0.596) (26,0.598) (27,0.572) (28,0.611) (29,0.616)
(30,0.587) (31,0.606) (32,0.614) (33,0.573) (34,0.600) (35,0.599)
(36,0.617) (37,0.590) (38,0.608) (39,0.600) (40,0.610) (41,0.578)
(42,0.603) (43,0.615) (44,0.618) (45,0.591) (46,0.589) (47,0.585)
};
\addlegendentry{Observed acceptance}

\addplot[
    red,
    dashed,
    thick
] coordinates {(-1,0.50) (48,0.50)};
\addlegendentry{Target Min 60\%}

\addplot[
    blue,
    dashed,
    thick
] coordinates {(-1,0.30) (48,0.30)};
\addlegendentry{Target Max 70\%}

\end{axis}
\end{tikzpicture}
\caption{Acceptance rate across configurations, showing stable APBNP negotiation behavior.}
\label{fig:48exp_line}
\end{wrapfigure}
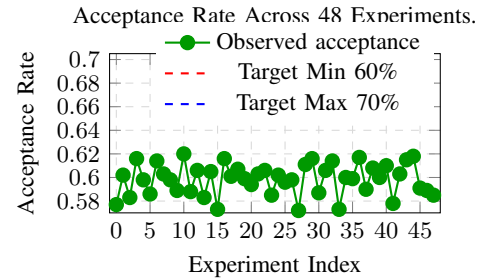
values without becoming stuck in infeasible scenarios. The negligible counter-offer rate suggests that most beneficial requests can be accepted directly, while the conservative budget usage reflects that prosumers do not fully deplete their privacy budgets, demonstrating the system's efficiency in resource consumption. From a deployment perspective, these 
\definecolor{XNegoBlue}{HTML}{1F77B4} 
\definecolor{StabilityRed}{HTML}{D62728} 
\definecolor{GridGray}{RGB}{200, 200, 200}
\begin{figure*}[htbp]
\centering
\begin{tikzpicture}

\begin{groupplot}[
    group style={
        group size=2 by 1,
        horizontal sep=1.5cm, 
    },
    width=0.48\textwidth,
    height=3 cm,
    tick label style={font=\small},
    label style={font=\small},
    title style={font=\small, align=center, yshift=5pt}, 
    major grid style={dashed, GridGray, line width=0.4pt},
]

\nextgroupplot[
    title={Mean Acceptance Rate Across Datasets},
    ylabel={Acceptance Rate},
    symbolic x coords={Oslo,Berlin,Rome,Paris,UCI,DE,FR,IT},
    xtick=data,
    xticklabel style={rotate=45, anchor=east, yshift=-2pt}, 
    ymin=0.975, ymax=1.005,
    ytick={0.98,0.99,1.00},
    ymajorgrids=true,
    ybar,
    height=4cm,
    bar width=14pt, 
    point meta=y,
    nodes near coords,
    nodes near coords style={
        font=\tiny, 
        text=white,
        inner sep=1.5pt, 
        anchor=north, 
    },
]

\addplot[
    dashed,
    line width=1.2pt, 
    StabilityRed,
    forget plot
] coordinates {(Oslo,0.98) (IT,0.98)} node[right, pos=1.0, font=\footnotesize, xshift=5pt, color=StabilityRed] {$\pm 0.02$ Band};

\addplot[
    dashed,
    line width=1.2pt,
    StabilityRed,
    forget plot
] coordinates {(Oslo,1.00) (IT,1.00)};

\addplot[
    fill=XNegoBlue!90, 
    draw=black,
    line width=0.5pt,
] coordinates {
    (Oslo,0.999)
    (Berlin,0.999)
    (Rome,0.998)
    (Paris,0.999)
    (UCI,0.999)
    (DE,0.999)
    (FR,0.998)
    (IT,0.999)
};

\nextgroupplot[
    title={Acceptance vs. Initial Privacy Budget},
    xlabel={Initial Privacy Budget $\epsilon_0$},
    ylabel={Acceptance Rate},
    xmin=0, xmax=10,
    xtick={0,2,4,5,6,8,10},
    ymin=0, ymax=1.05,
    ytick={0,0.2,0.4,0.6,0.8,1.0},
    xmajorgrids=true,
    ymajorgrids=true,
    height=3cm
]

\addplot[draw=none, fill=StabilityRed!20, opacity=0.8]
    coordinates {(0,0) (5,0) (5,1.05) (0,1.05)} \closedcycle;
\node at (2.5, 0.1) [font=\footnotesize, text=StabilityRed!80!black] {Scarcity-Driven Rejection};

\addplot[draw=none, fill=green!20, opacity=0.8]
    coordinates {(5,0) (10,0) (10,1.05) (5,1.05)} \closedcycle;
\node at (7.5, 0.1) [font=\footnotesize, text=green!50!black] {Near-Perfect Acceptance};

\addplot[
    line width=1.5pt, 
    XNegoBlue,
    domain=0:10,
    samples=200,
    smooth
] {1/(1 + exp(-1.5*(x-5)))};

\draw[dashed, gray!70] (axis cs:5, 0) -- (axis cs:5, 1.05); 
\node at (axis cs:5, 1.03) [anchor=south, font=\small] {$\epsilon_0=5$ (Transition Point)};

\end{groupplot}
\end{tikzpicture}

\caption{Acceptance stability and privacy-regime transitions in \textsc{X-NegoBox}.
(\textbf{Left}) Mean acceptance remains within a narrow band ($\pm0.02$) across datasets.
(\textbf{Right}) Increasing initial privacy budgets transitions APBNP from scarcity-driven
rejection to near-perfect acceptance.}
\label{fig:privacy_sweep}
\end{figure*}
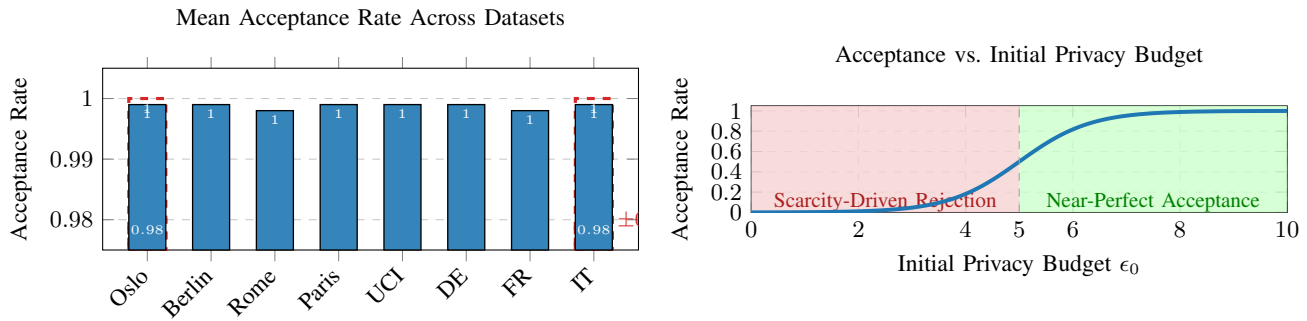
findings suggest that X-NegoBox can sustain long-term operation without rapidly exhausting prosumers’ privacy budgets, while still enabling a high fraction of socially useful contracts. As shown in Fig.~\ref{fig:48exp_line}, acceptance rates remain tightly concentrated around 60\%, 
confirming stable negotiation behavior across patterns and datasets. Fig.~\ref{fig:privacy_sweep} provides a structured view of how acceptance behaves jointly 
across datasets and policy patterns, reinforcing the observation of dataset-independent stability. The transition observed in Fig.~\ref{fig:privacy_sweep} empirically validates the theoretically 
predicted privacy-regime behavior: relaxations in $\epsilon_0$ enlarge the feasible negotiation region.

\textbf{Privacy-Budget Sweep:}
Figure~\ref{fig:sweep_plot} shows acceptance, rejection, and counter-offer rates
as functions of the initial privacy budget $\varepsilon_0 \in \{1,\dots,10\}$,
revealing four operating regimes.
In the \textbf{scarce-privacy regime} ($\varepsilon_0 \le 2$), acceptance ranges
from 55\% to 74\%, with frequent rejections required to preserve privacy.
The \textbf{transitional regime} ($2 < \varepsilon_0 < 4$) exhibits a smooth
increase in acceptance as additional budget becomes available.
In the \textbf{stable regime} ($4 \le \varepsilon_0 \le 7$), acceptance saturates
around 99\% with negligible rejection. Finally, the \textbf{surplus regime} ($\varepsilon_0 \ge 8$) corresponds to a
setting in which privacy is no longer a binding constraint: acceptance reaches
$1.00$ and remains constant, indicating that further budget increases do not
affect negotiation outcomes. This behavior reflects diminishing returns in privacy-budget allocation.
At low $\varepsilon_0$, privacy dominates feasibility, and small budget increases
substantially improve acceptance. Beyond moderate budgets
(approximately $\varepsilon_0 \in [4,5]$), this constraint weakens and other
factors, such as data sensitivity, declared purpose, and remaining cumulative
budget, become decisive. From a system-design perspective, these results indicate that excessively large
initial privacy budgets are unnecessary: comparable acceptance performance can be
achieved with moderate privacy flexibility, preserving long-term privacy without
sacrificing negotiation success.

\textbf{Discussion, Limitations, and Future Work:}
Results show that X-NegoBox effectively balances utility and privacy, achieving stable acceptance without tuning and high acceptance (87.49\%) with minimal counter-offs. It manages budgets conservatively and adapts to trust and sensitivity variations. Limitations include simulation-based evaluation, static purpose scores $P(p)$, lack of network-level TSS validation, and absence of user studies. Future work will explore real-world deployment, adaptive purpose modeling, negotiation-aware cryptography, and user-centric evaluation. \footnote{\textbf{Ethics and Reproducibility:}
X-NegoBox follows privacy-by-design, keeping raw data local and releasing only DP outputs. Experiments use public/synthetic data and are reproducible (\url{https://github.com/Poushali96/X-NEGOBOX}).This work was supported by dScience, University of Oslo, and the UiO Energy Convergence Environment PriTEM. The authors gratefully acknowledge Prof. Olaf Owe for his support.}
\section{Conclusion}
This paper presents \textit{X-NegoBox}, an explainable and adaptive privacy-budget negotiation framework for secure peer-to-peer energy data exchange. By combining local execution with APBNP and X-Contract, it replaces static policies with context-aware negotiation. Experiments across diverse datasets show stable acceptance, sustained data sharing, and improved transparency, supporting privacy-aware decentralized energy systems. Future work includes negotiation-aware TSS, incentive design, and large-scale deployment.


    





\bibliographystyle{IEEEtran}
\bibliography{name}

\end{document}